\documentclass[apjl]{emulateapj}

\newcommand{\arcm}{\ifmmode {' }\else $' $\fi}
\newcommand{\arcs}{\ifmmode {'' }\else $'' $\fi}
\slugcomment{}

\shortauthors{Rhode, Zepf, \& Santos} \shorttitle{Globular Clusters and Galaxy
Formation}

\begin{document}

\title{Metal-Poor Globular Clusters and the Formation of Their Host Galaxies}

\author{Katherine L. Rhode\altaffilmark{1,2,3}}

\author{Stephen E. Zepf\altaffilmark{4}}

\author{Michael R. Santos\altaffilmark{5}}

\altaffiltext{1}{Astronomy Department, Wesleyan University, Middletown, CT 06459; kathy@astro.wesleyan.edu}
\altaffiltext{2}{Department of Astronomy, Yale University, New Haven, CT 06520}
\altaffiltext{3}{NSF Astronomy \& Astrophysics Postdoctoral Fellow}
\altaffiltext{4}{Department of Physics \& Astronomy, Michigan State University,
  East Lansing, MI 48824; zepf@pa.msu.edu}
\altaffiltext{5}{Institute of Astronomy, University of Cambridge, Cambridge CB3 0HA, UK; mrs@ast.cam.ac.uk}

\begin{abstract}
We have determined the total numbers and specific frequencies of blue,
metal-poor globular clusters (GCs) in eight spiral and early-type
galaxies. These data, along with five measurements from the
literature, show a trend of increasing blue GC specific frequency with
increasing mass of the host galaxy.  The increase is not accounted for
in a simple galaxy formation model in which ellipticals and their GC
systems are formed by the merger of typical spiral galaxies.  The data
appear broadly consistent with hierarchical formation scenarios in
which metal-poor GCs are formed over a finite period in the early
Universe during the initial stages of galaxy assembly.  In this
picture, the observed trend is related to biasing, in the sense that
the more massive galaxies of today began assembling earlier and
therefore formed relatively more GCs during this early epoch of
metal-poor GC formation.  We discuss how comparisons of the observed
specific frequency of metal-poor GCs with model calculations can
constrain the formation redshift of these objects.
\end{abstract}


\section{Introduction}

The precise way that galaxies like our own Milky Way form is one of
the central issues in astrophysics and cosmology today.  One method
for investigating a galaxy's origins is to study its stellar
populations and use their properties to piece together a coherent
picture of the different stages of the galaxy's history.  Globular
clusters (GCs) make particularly useful probes of galaxy formation
because these luminous, compact star clusters are easy to see at large
distances and appear in all types of galaxies, and the oldest GCs have
ages approaching the age of the Universe (e.g., Ashman \& Zepf 1998,
Chaboyer et al.\ 1998).  Furthermore, GCs are thought to form during
periods of intense star formation and therefore trace the major
star-forming episodes in a galaxy's past.  Carefully quantifying the
observable properties of the entire system of GCs in a galaxy leads to
important insights into that galaxy's origins.

With this rationale in mind, we initiated a ground-based imaging
survey of the GC systems of several giant galaxies (Rhode \& Zepf
2001, 2003, 2004; hereafter, RZ01, RZ03, RZ04).  The survey was
designed to produce reliable determinations of the total number of GCs
in each galaxy and to quantify the spatial and color distributions of
the GCs out to large galactocentric radius.  We use large-format CCDs
to observe the entire radial extent of the GC systems.  Furthermore,
our survey methods (e.g., three-color photometry) facilitate the
separation of real GCs from contaminating foreground and background
sources.  As a result, we have significantly reduced the levels of
contamination in our samples
and reduced the errors on our calculated GC specific frequencies by a
factor of 2$-$3 compared to past studies (RZ01, RZ03, RZ04).

The survey's main purpose is to help answer questions about how
galaxies and their GC systems form.  In particular, we aimed to test
the predictions of models for the origins of ellipticals.  Two
important properties of elliptical galaxy GC systems are that they are
typically more populous than those of spiral galaxies and that they
often show two peaks in their color or metallicity distributions
(Ashman \& Zepf 1998 and references therein).  Ashman \& Zepf (1992;
hereafter AZ92) predicted this bimodality when they proposed that
ellipticals and their GC systems are created in mergers of disk
galaxies.  In this picture, a metal-poor (blue) GC population in the
elliptical comes from the progenitor spiral galaxies, and a metal-rich
(red) population is formed during the merger.  A consequence of this
scenario is that the relative numbers of metal-poor GCs should be
similar in spiral and elliptical galaxies.  This can only be tested
with a careful accounting of the total numbers and specific
frequencies of GCs (both blue and red) in a sample of early and
late-type giant galaxies.  Such data are also useful for testing other
models proposed to account for the observed bimodal metallicity
distributions (e.g., Forbes, Brodie \& Grillmair 1997, C\^ot\'e,
Marzke, \& West 1998).

More broadly, the low metallicities (the mean [Fe/H] for the
metal-poor Milky Way GCs is $-$1.6; Armandroff \& Zinn 1988) and
extended spatial distributions (RZ01, RZ04) of metal-poor GCs suggest
that they were some of the first structures to form in the Universe.
Therefore we can use what we learn about metal-poor GC systems to
probe structure formation at early epochs.  In hierarchical
cosmological models, galaxy halos that are more massive and are
located within dense environments will tend to have a higher
proportion of their mass accumulated by a given redshift compared to
less-massive galaxies in poorer environments (e.g., Sheth 2003).  The
assumption of a preferred formation epoch and constant formation
efficiency for metal-poor GCs then leads to the prediction that
higher-mass galaxies will have larger relative numbers of metal-poor
GCs.  Furthermore, the exact spatial distribution of the metal-poor
GCs out to large radius constrains the formation redshift and the
masses of the structures within which they formed (Santos 2003 and
references therein).

We have recently completed the analysis for three more galaxies from
the survey (Rhode \& Zepf 2005) bringing our total sample to eight
spiral, elliptical, and S0 galaxies.  We can combine these results
with those from other studies to address some of the questions that
motivated the survey and to investigate the possible cosmological bias
of metal-poor GCs.  In this letter, we present our results for the
specific frequency of metal-poor GCs and discuss what they might imply
about how and when galaxies and their GC systems formed.

\section{Specific Frequency of Metal-Poor Globular Clusters}
\label{section:Tblue}

\subsection{Survey Data and Analysis}
\label{section:analysis}

Details of our data analysis methods are given in a series of previous
papers (RZ01, RZ03, RZ04).  Briefly, to estimate the number of
metal-poor GCs in a target galaxy, we begin by imaging the galaxy in
three broadband optical filters ($BVR$ or $BVI$).  GC candidates are
selected by their radial profiles (GCs appear unresolved from the
ground at the 10$-$20~Mpc distances of our targets), apparent
magnitudes, and colors.  The spatial distribution of GC candidates for
each galaxy is then constructed and corrected for incompleteness due
to the magnitude depth of the images, missing spatial coverage around
the galaxy, and contamination from non-GCs.  Integrating this
distribution out to the radius at which the GC surface density drops
to zero yields an estimate of the total number of GCs in the system.

As mentioned earlier, the broadband color distributions of many giant
galaxies often show two peaks, indicating the presence of two
populations of GCs (e.g., Zepf \& Ashman 1993; Gebhardt \&
Kissler-Patig 1999; Kundu \& Whitmore 2001; RZ01; RZ04).  The color
difference represents a difference in the metallicity of the two
populations in nearly all cases, because the GC systems are typically
more than a few Gyr old (Zepf \& Ashman 1993; Ashman \& Zepf 1998).
In a few well-studied galaxies, the spatial distributions and
kinematics of the blue (metal-poor) and red (metal-rich) GCs have been
shown to differ from each other, supporting the interpretation that
these are distinct populations created in different star formation
episodes (e.g., Zinn 1985, Zepf et al.\ 2000).  To calculate a total
number of blue GCs in our galaxies, we use the KMM mixture-modeling
code (Ashman, Bird, \& Zepf 1994), which (in addition to testing
whether a single or multiple Gaussian function provide a better fit to
a distribution) locates the peaks and dispersions of the Gaussians
that best fit a distribution and provides an estimate of the
fractional contribution of each population.  The code is run on a
sample of GC candidates for which our detection is complete in all
three filters.

For the early-type galaxies in the sample, the separation between the
blue and red GCs occurs at $B-R$~$\sim$1.2$-$1.3, and 60$-$70\% of the
GCs are blue (RZ04).  The spiral galaxies in the study often have
poorer statistics, making definitive separation of the candidates into
blue and red populations difficult or impossible.  In such cases we
counted as metal-poor the GCs bluer than $B-R$~$\sim$~1.23, the
typical location of the separation in the color distributions of the
early-type galaxies (RZ04).  The gap between the metal-poor and
metal-rich GCs in the Milky Way occurs at $B-R$~$\sim$~1.25
\citep{harris96},
indicating this is a reasonable choice.  The estimated
blue/red proportions of GCs in the spiral galaxies were not sensitive
to the exact location of the $B-R$ separation; using 1.23 or 1.25
produced the same results within the errors.  The total number of blue
GCs was computed for each galaxy by multiplying the estimated blue
fraction by the total number in the system.

\subsection{Measurements from the Literature}
\label{section:literature}

To supplement our data, we searched the literature for well-determined
measurements of the numbers of blue GCs in other giant galaxies.  We
used the following criteria for inclusion of a particular measurement
here: at least 50\% of the estimated radial extent of the GC system
must have been observed; data must have been acquired in at least two
filters, so the GC color distribution could to be quantified; and
estimates of the total number of GCs and the blue fraction were given,
or could be derived in a fairly straightforward way from the published
data.  Finally, the measurement was included only if its 1-$\sigma$
error was $_<\atop{^\sim}$40\%.  A comprehensive search of refereed
journal papers yielded measurements that met these criteria for
three elliptical galaxies. We also include estimates of the number of
blue GCs in the two most well-studied giant galaxies, the Milky Way
and M31, in our analysis.  The data for all the galaxies and the
references from which they are derived are given in
Table~\ref{table:tblue}.

\subsection{Results}

In order to compare the blue GC populations of the galaxies, we have
calculated the mass-normalized number of blue GCs, or $T_{\rm~blue}$,
for each.  This is a variation of the $T$ parameter introduced by
\citet{za93} and is defined as:

\begin{equation}
T_{\rm blue} \equiv \frac{N_{GC}(\rm blue)}{M_G/10^9\ {\rm M_{\sun}}}
\end{equation}

\noindent where $N_{GC}(\rm blue)$ is the number of blue GCs and $M_G$
is the stellar mass of the host galaxy.  Normalizing $N_{GC}$ by
galaxy mass facilitates the comparison of the GC systems of galaxies
with a range of morphological types.  Normalizing instead by galaxy
luminosity within a given waveband --- e.g., as in the specific
frequency $S_N$ of \citet{hvdb81} --- makes it difficult to directly
compare numbers of GCs in spirals and ellipticals with very different
stellar populations.


\begin{figure}
\plotone{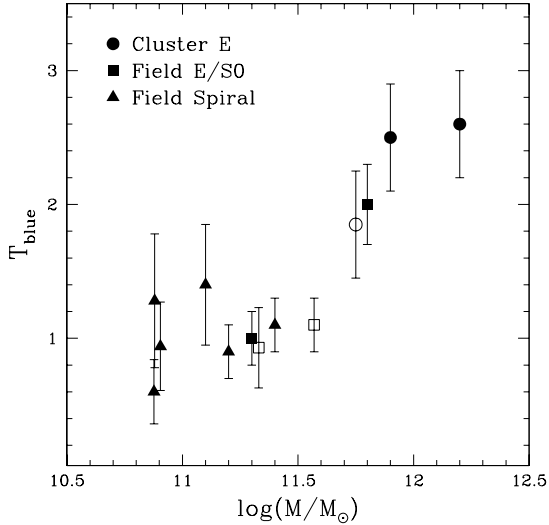}
\caption{Mass-normalized numbers of blue GCs for 13 giant galaxies
versus the log of the galaxy stellar mass.  Circles are cluster
ellipticals, squares are early-type field galaxies, and triangles are
field spiral galaxies.  Filled symbols are used for our data and the
Local Group galaxies; open symbols are data from other studies.
\protect\label{fig:tblue mass}}
\end{figure}

Figure~\ref{fig:tblue mass} shows $T_{\rm~blue}$ for 13 giant
galaxies: eight from our survey and five from other work.  To
calculate galaxy masses, we assume mass-to-light ratios ($M/L_V$)
following those of \citet{za93}: $M/L_V$~$=$~10 for ellipticals,
$M/L_V$~$=$~7.6 for S0 galaxies, and $M/L_V$ between 5.0 and 6.1 for
spirals, depending on Hubble type.  The relevant quantity for
comparing galaxies across Hubble types is the relative mass-to-light
ratio as a function of type, and analyses of both stellar populations
and dynamics yield relative mass-to-light ratios similar to those
adopted here (e.g., Ashman \& Zepf 1998).  Kauffmann et al.\ (2003)
analyzed Sloan Digital Sky Survey data and also found similar or
slightly larger mass-to-light ratio differences among galaxy types.

$T_{\rm blue}$ is shown in the figure with different symbols to
indicate galaxy type and environment: circles for cluster ellipticals,
squares for E/S0 galaxies in the field or in groups, and triangles for
field spiral galaxies.  The numbers of blue GCs for the survey
galaxies were computed as outlined in Section~\ref{section:analysis}.
The uncertainties on our $T_{\rm blue}$ values include Poisson errors
on the number of GCs and contaminating objects, errors related to the
magnitude completeness of the data, and errors on the galaxy
magnitudes and distances (see RZ04 for example error calculations).
We computed $T_{\rm blue}$ for galaxies in other studies by combining
the published $N_{GC}$ and blue fraction with the galaxy absolute
magnitude and appropriate $M/L_V$.  The error bars are based on the
published errors on $N_{GC}$ along with uncertainties on the galaxy
magnitude and distance.  Quantities shown in Figure~\ref{fig:tblue
mass} are listed in Table~\ref{table:tblue}.


\begin{deluxetable}{lllcll}
\tablecaption{Properties of Galaxies in Figure~\ref{fig:tblue mass}}
\tablehead{\colhead{Name}&\colhead{Type}&\colhead{$M_V^T$}
  &\colhead{Mass} &\colhead{$T_{\rm blue}$}&\colhead{Reference}\\
\colhead{} & \colhead{} & \colhead{} & \colhead{[log($\rm M/M_{\sun}$)]} & \colhead{} & \colhead{}}
\startdata
NGC4472 & E2  & $-$23.1 & 12.2   & 2.6$\pm$0.4 & 6\\
NGC4406 & E3  & $-$22.3 & 11.9   & 2.5$\pm$0.4 & 8\\
NGC4374 & E1  & $-$22.1 & 11.8   & 1.9$\pm$0.4 & 4\\
NGC4594 & S0  & $-$22.4 & 11.8   & 2.0$\pm$0.3 & 8\\
NGC5128 & E0  & $-$21.9 & 11.6    & 1.1$\pm$0.2 & 5\\
NGC3379 & E1  & $-$20.9 & 11.3   & 1.0$\pm$0.2 & 8\\
NGC1052 & E4  & $-$21.0 & 11.3    & 0.9$\pm$0.3 & 3\\
NGC7814 & Sab & $-$20.4 & 10.9   & 1.3$\pm$0.5 & 7\\
NGC4157 & Sb  & $-$20.4 & 10.9   & 0.6$\pm$0.2 & 9\\
NGC2683 & Sb  & $-$20.5 & 10.9   & 0.9$\pm$0.3 & 9\\
M31 & Sb & $-$21.8     & 11.4    & 1.1$\pm$0.2 & 1, 2\\
Galaxy & Sbc & $-$21.3 & 11.2    & 0.9$\pm$0.2 & 1, 10\\
NGC3556 & Sc  & $-$21.2 & 11.1   & 1.4$\pm$0.5 & 9
\enddata
\tablecomments{NGC4472, NGC4406 and NGC4374 are Virgo cluster
  galaxies; all others are in the field or in groups.  $M_V$ was derived
  as follows: for NGC4472, by combining $V$ from RZ01 with $m-M$ from
  Whitmore et al. 1995; for NGC4406 and NGC4374, combining $V$ from RC3
  (deVaucouleurs et al. 1991) with $m-M$ from Whitmore et al. 1995;
  for NGC3556, from RC3; for NGC5128, combining $V$ from Dufour et
  al. 1979 with $m-M$ from Israel 1998; for the Milky Way and M31,
  from Ashman \& Zepf 1998; and for all other galaxies, combining $V$
  from RC3 with $m-M$ from Tonry et al. 2001. 
}
\tablerefs{(1) Ashman \& Zepf 1998; (2) Barmby et al.\ 2000; (3)
  Forbes, Georgakakis, \& Brodie 2001; (4) Gomez \& Richtler 2004; (5)
  Harris, Harris, \& Geisler 2004; (6) Rhode \& Zepf 2001; (7) Rhode
  \& Zepf 2003; (8) Rhode \& Zepf 2004; (9) Rhode \& Zepf 2005; (10)
  Zinn 1985.}
\protect\label{table:tblue}
\end{deluxetable}

The data in Figure~\ref{fig:tblue mass} show an overall trend of
increasing $T_{\rm blue}$ with increasing galaxy mass.  The two most
massive galaxies included --- the Virgo cluster ellipticals NGC~4406
and NGC~4472 --- have 2$-$2.5 times the $T_{\rm blue}$ values of the
lower-mass field galaxies on the left side of the plot.  A factor that
may contribute to this trend bears discussion before we attempt to
interpret the data.  Following \citet{za93}, we used a constant
$M/L_V$
to calculate $M_G$ for the elliptical galaxies.  In fact, the stellar
mass-to-light ratios of ellipticals almost certainly vary with galaxy
luminosity, because more luminous ellipticals are redder (this is the
well-known color-magnitude relation). The redder stellar populations
of more luminous ellipticals invariably have higher $M/L_V$ in any
standard stellar populations model.

The detailed dependence of $M/L_V$ with luminosity goes as
$M/L_V$~$\propto$~$L^{~\sim0.07}$ if the color-magnitude relation is
due to metallicity effects (e.g., Dressler et al.\ 1987), or as
$M/L_V$~$\propto$~$L^{~\sim0.10}$ if most of the effect is due to age
differences (Zepf \& Silk 1996 and references therein).  If we assume
the latter variation of $L^{0.10}$, then $T_{\rm~blue}$ for the
luminous Virgo ellipticals would be reduced by a factor of $\sim$1.2
relative to the value for the less massive elliptical, NGC~3379.
Since $T_{\rm blue}$ for the massive Es is more than twice that of the
lower-mass galaxies in the figure, possible $M/L_V$ variations can
only account for about one-third of the observed difference in $T_{\rm
blue}$.  Thus the trend of increasing $T_{\rm blue}$ with increasing
host galaxy mass is significant and is likely to have important
implications for galaxy and GC formation.

\subsection{Interpretation and Implications for Galaxy Formation}
\label{section:implications}

The prediction of AZ92 that spiral and elliptical galaxies should have
similar relative numbers of metal-poor GCs 
can be directly tested using the data in Figure~\ref{fig:tblue mass}.
$T_{\rm blue}$ for the six spiral galaxies has a weighted mean value
of 1.2$\pm$0.1.  $T_{\rm blue}$ for the massive Virgo cluster
ellipticals is twice that of the spirals.  We noted earlier that, even
considering the possible effects of stellar populations variations
among ellipticals, the difference in $T_{\rm blue}$ between the
spirals and cluster ellipticals is significant and indicates that, in
terms of the total numbers, combining the metal-poor GC populations in
a typical spiral galaxy is not sufficient to produce the metal-poor GC
population of a luminous elliptical galaxy.  A different --- or
additional --- formation mechanism besides the straightforward merging
of AZ92 must be responsible for creating the populous metal-poor GC
populations of luminous ellipticals in high-density environments like
those plotted with filled circles in Figure~1.

More generally, the data in Figure~1 show a rough trend of larger
  relative numbers of metal-poor GCs with increasing galaxy mass.
  Such a trend could be explained by a scenario in which galaxies are
  formed hierarchically and metal-poor GC formation occurs over a
  finite period in the early Universe.  As we noted earlier, in
  hierarchical pictures more massive galaxies have a greater fraction
  of their mass assembled by a given redshift than less massive
  galaxies.  If GC formation was favored at a specific epoch early in
  the Universe's history, then more massive galaxies would
  consequently have a greater fraction of their mass in GCs.  Because
  the GCs that formed in this early epoch would be metal-poor, the
  result would be that massive ellipticals like NGC~4472 and NGC~4406
  in the Virgo cluster would today have higher $T_{\rm blue}$ values
  than less massive galaxies.  In this type of picture, lower-mass
  ellipticals like NGC~3379 start their formation later and therefore
  form a greater fraction of their stars at a later time, when
  metal-poor GC formation is no longer favored.  Thus they have lower
  mass-normalized specific frequencies of metal-poor GCs.

\citet{santos03} suggested that the type of scenario just outlined is
indeed responsible for differences in the blue GC specific frequencies
of galaxies of different masses and environments.  In the proposed
picture, metal-poor GCs formed at high redshift when gas-rich
protogalactic building blocks merged to create larger structures.
This structure formation is suppressed temporarily by reionization.
Stellar evolution continues to enrich the intergalactic medium in the
intervening time, so that once structure formation resumes, any GCs
that are created are comparatively metal-rich.  Because in this
picture metal-poor GC formation occurs over a finite period, and the
redshift at which a given galaxy began its assembly varies according
to the final mass of the galaxy and its location in the Universe,
differences in $T_{\rm blue}$ are produced for galaxies of different
masses and environments.

Detailed modeling of how the relative numbers of blue GCs should
change with galaxy mass and environment for a given GC formation epoch
has yet to be accomplished.  We have, however, compared our data with
a preliminary calculation from \citet{santos03} in which blue GCs form
before $z~=~10$.  The $T_{\rm blue}$ differences we observe are no
larger than what would be expected if metal-poor GCs are formed prior
to this redshift.  Furthermore, the observed differences appear to be
smaller than what is expected if metal-poor GCs form at redshifts much
greater than 10.  More theoretical work is required before detailed
comparisons between data and model predictions can be made.  But the
idea of a preferred early formation epoch for the metal-poor GCs,
combined with biased hierarchical galaxy formation, is appealing as a
possible explanation for the apparent trend in the data.

As described in Santos~(2003), the observed spatial distributions of
metal-poor GCs can help test the idea of a biased, hierarchical origin
for galaxies and their GC systems.  This is because structure that
forms before a fixed redshift threshold is more centrally-concentrated
than the dark matter distribution of the final galaxy.  Likewise GCs
formed within higher-mass protogalactic fragments will be more
concentrated toward the center of a galaxy's dark matter halo than
those formed in lower-mass fragments.  Testing whether specific
hierarchical models work requires predictions for the metal-poor GC
spatial distributions of galaxies with a range of masses and
environments for comparison with our observed radial distributions
(RZ01, RZ03, RZ04).

Perhaps relevant to this discussion is a recent study by Strader,
Brodie, \& Forbes (2004) that finds a weak correlation between the
location of the blue GC peak in a sample of giant galaxies and the
galaxies' luminosities.  They state that the presence of the
correlation means that the metal-poor GCs ``knew'' about the galaxy to
which they would eventually belong and may have been located within
that galaxy's dark matter halo when they formed.  We suggest that
within the hierarchical framework outlined above, one might expect
that because the assembly process begins earlier for luminous galaxies
in denser environments, chemical enrichment (for similar reasons)
proceeds more quickly and efficiently in these locations, so that
metal-poor GCs formed in these galaxies are slightly more metal-rich
than their counterparts in less luminous field galaxies.  This could
produce a shallow positive correlation between the mean metallicity of
the metal-poor GCs and galaxy total luminosity.  Further study of how
the blue GC systems of galaxies in the field compare to those of
galaxies in dense environments would help to explore this possible
explanation.

We have emphasized here the links that the metal-poor GCs have with
galaxy formation, but the metal-rich GCs are perhaps equally important
as signposts of what has occurred over the course of a galaxy's
history.  In the type of hierarchical picture described, the galaxy
assembly process begun at high redshift would be expected to continue
to the current epoch, with the giant elliptical and spiral galaxies of
today being the product of many major and minor mergers. Formation of
metal-rich GCs was predicted and has been observed to occur during
these mergers, provided they are gas-rich (e.g., AZ92, Schweizer 1987,
and many others).  The proportion of red and blue GCs, as well as how
broad the metallicity range of the final metal-rich GC peak is, would
then depend on the exact merger and assembly history of a particular
galaxy.  This idea is borne out in a general sense by results from
semi-analytic galaxy formation models (e.g., Beasley et al.\ 2002),
which give rise to broadband GC color distributions that have a large
range in appearance and that reflect the wide variation of possible
galaxy accretion histories.  The GC color distributions we observe
likewise exhibit a range of morphologies (RZ01, RZ04).  Our survey
data may also suggest that $T_{\rm red}$, the number of red GCs
normalized by galaxy stellar mass, increases with galaxy mass (RZ04).
This might be accounted for in hierarchical merger models if a larger
number of gas-rich mergers contribute to the assembly of more massive
ellipticals located in richer environments.



\acknowledgments

KLR is supported by an NSF Astronomy and Astrophysics Postdoctoral
Fellowship under award AST-0302095.  SEZ is supported by NSF award
AST-0406891. MRS is supported by an NSF MPS-DRF under award
AST-0302148.



%
%

\end{document}